\documentstyle[preprint,aps]{revtex}
\begin{document}
\draft
\preprint{CEBAF TH-95-10, CU 1120}
%
%   REVTEX Mode
%

\title{A Comment on General Formulae for Polarization Observables \\
in Deuteron Electrodisintegration and Linear Relations}
\author{V. Dmitra\v sinovi\' c}
\address{
Physics Department, University of Colorado, \\
Nuclear Physics Lab, P.O. Box 446, Boulder, CO 80309-0446}
\author{Franz Gross}
\address{Department of Physics,
College of William and Mary, Williamsburg, VA 23185 \\
and Physics Division, MS12H2, Continuous Electron Beam Accelerator Facility,\\
12000 Jefferson Ave, Newport News, VA 23606}
%\receipt{}
\maketitle
\begin{abstract}
We establish a simple, explicit relation between the formalisms employed
in the treatments of polarization observables in deuteron two-body
electrodisintegration published by Arenh\"ovel, Leidemann, and
Tomusiak in Few-Body Systems {\bf 15}, 109 (1993)
and the results of the present authors published in
Phys.~Rev.~C {\bf 40}, 2479 (1989). We comment on the overlap between the
two sets of results.
\end{abstract}
%\pacs{PACS numbers: 25.30.-c, 24.70.+s, 13.60.-r}

\section*{}

In a recent issue of this journal an article \cite{ar93} by Arenh\"ovel,
Leidemann and Tomusiak (ALT) on ``General Formulae for Polarization
Observables in Deuteron Electrodisintegration and Linear Relations" has
appeared.  Four years earlier \cite{dg89} we published a comprehensive
treatment of polarization observables in this reaction (DG),
and since the ALT paper makes no reference to our work
we feel obliged to comment on these two papers, and to discuss the
relationship between these two approaches. In this comment we will establish a
simple and explicit relation between the transition amplitudes in the two
approaches, whereupon all of our results \cite{dg89} become immediately
applicable to the ALT formalism.

Before we compare these two papers in detail, we review the arguments which
determine the number of real observables which can be measured in deuteron
electrodisintegration.  The total number of spin variables in this reaction
are $3\times3\times2\times2= 36$, but because of the parity constraint, only
half of these complex amplitudes, 18, are independent.  The number of real
bilinear products which can be formed from these 18 complex amplitudes is
$18\times18=324$ (since $A^*B$ and $B^*A$ are equivalent to two real
functions).  However, since there are ``only'' 18 independent complex
amplitudes, and since the overall phase can never be determined, all of
these 324 observables depend on products of only 35 independent real
functions.  The problem of completely measuring deuteron
electrodisintegration reduces to the problem of designing a program of
measurements from which the 35  independent
real functions can unambigously extracted from combinations of the 324
bilinear
products measured in actual experiments.  Clearly not {\it all\/} possible
measurements are needed for a complete determination, and as more and more
measurements are added to the data base, greater and greater care must be
taken to find new measurements which give truly independent information.

Because of this redundancy, in DG we discussed all possible spin observables
which can be measured in the reaction $d(e,e'N_1)N_2$, where nucleon $N_2$
is not observed, and therefore its polarization is not detected.  Hence we
limited ourselves to observables in which the polarization of the virtual
photon, the deuteron
target, and {\it one\/} outgoing nucleon are measured, either singlely or
in all possible combinations.  Choosing a hybrid transversity basis we were
able to obtain a comparatively simple result, and demonstrated that 162
bilinear products of amplitudes can be measured by looking at reactions
where $N_1=p$.  Adding the cases where $N_1=n$ gives another 162 bilinear
products of amplitudes, but only 80 of these are new (see
below).  We did not discuss measurements in which the polarization of  both
of the outgoing nucleons are measured (which requires
$d(e,e'\,\vec{n}\,\vec{p})$  measurements)
but the remaining 82 products could be measured in this way.  We
also showed that the 162 observables accessible to $N_1=p$ measurements
could {\it not\/} completely determine the 35 independent real quantities,
{\it even though 162 is far greater that 35\/}.  At least one neutron
polarization measurement must be made before all 35 independent real
quantities can be extracted, but one such additional measurement is
sufficient, in principle, to complete the program.

The ALT paper is an extension of an earlier paper \cite{ar88} on
complete classification of all polarization experiments in
deuteron photodisintegration. They extend their photodisintegration
formalism   to include longitudinal polarization of the virtual photon, and
the number of amplitudes is accordingly increased from 12 to 18.  They
discuss {\it all\/} possible polarization measurements, including those
which can be obtained from $d(e,e'\,\vec{n}\,\vec{p})$, and hence should
obtain all of the 324 bilinear  products.  However, their formalism
generates twice this many (648), and  they spend some time showing how the
parity constraint generates the necessary 324 linear relations between these
648 bilinear products, all of which are nonzero in their formalism.
Unfortunately, the linear relations between the 648 amplitudes
make it difficult to see which measurements are sufficient to extract the
35 truly independent real functions needed to completely
determine all deuteron electrodisintegration observables.  At the end of
their
paper they say that, in order to fully determine all observables in
deuteron electrodisintegration, ``one cannot totally avoid'' measuring two
observables  in which the polarizations of {\it both\/} outgoing nucleons
are measured.   This conclusion contradicts the result of DG, where we
showed that {\it no\/} such measurements are {\it required\/} (although it
might turn out that  a particular separation strategy might make use of such
measurements).  We will compare the results
of these two papers in more detail shortly.

Before turning to the details of this discussion, it might help the
reader to look at these two papers within the historical context.
The ALT paper is the latest in a long series\cite{ar88,ar79,alt88} that
can be  traced back to J.J. De Swart's founding paper from 1959
\cite{des59}, which  relied on the nonrelativistic spin polarization
formalism developed by  Ashkin and Wolfenstein \cite{aw52}.

Our approach has a similarly long
lineage dating back to the relativistic spin polarization (helicity)
formalism  of Jacob and Wick (JW) \cite{jw59} who, among other things,
established a  link with the nonrelativistic formalisms. This method allows
a simple  exploration of parity and other symmetries, as well as complete
separations of  amplitudes.
The first application of the JW formalism to
the problem of deuteron photo- and electrodisintegration was made by
LeBellac,  Renard and Tran Thanh Van in a series
\cite{lrt164,lrt264,rtb165,rtb265,rtb365} of formal and practical papers
in the mid-sixties. At that time the
helicity formalism was still
sufficiently new to warrant a comparison with the older nonrelativistic
formalism. The relation between the two, in the form of formulae for
multipoles, was explicitly spelled out in Appendix B
of Ref. \cite{lrt164} for the photodisintegration amplitudes and in
section 4.2
of Ref. \cite{rtb165}, as well as in section 5 of Ref. \cite{rtb265} for
the  electrodisintegration amplitudes.  Although the helicity formalism has
its complexities, it does lead to final results with patterns simple enough
to allow for a comparatively simple discussion of separation strategies.

We now turn to a detailed comparison of the ALT and DG papers.  Our approach
(DG) begins with the use of helicity amplitudes with parity transformation
properties summarized by the following relation
\begin{mathletters}
\begin{eqnarray}
\langle \lambda_{p}~\lambda_{n}\vert J_{\lambda_{\gamma}}\vert \lambda_{D}
\rangle
&\equiv& \langle \lambda_{p}~\lambda_{n}\vert J\cdot
\epsilon_{\lambda_{\gamma}} \vert \lambda_{D}\rangle \\
\langle \lambda_{p}~\lambda_{n}\vert J_{\lambda_{\gamma}}\vert \lambda_{D}
\rangle &=& (-1)^{(\lambda_{p} - \lambda_{n})-
(\lambda_{\gamma} - \lambda_{D})}
\langle -\lambda_{p}~-\lambda_{n}\vert J_{-\lambda_{\gamma}}
\vert -\lambda_{D}\rangle~, \label{parity}
\end{eqnarray}
\end{mathletters}
\noindent where $J \cdot \epsilon_{\lambda} = J_{\mu} \;
\epsilon^{\mu}_{\lambda}$, using
the Bjorken and Drell metric \cite{bd65}, and the initial state consists
of a virtual photon with helicity $\lambda_\gamma$ and a deuteron (particle
No.~2 in the sense of Jacob and Wick \cite{jw59}) with helicity
$\lambda_D$, and the
final state consists of an outgoing proton with helicity $\lambda_p$ and
neutron (particle No.~2) with helicity $\lambda_n$.
The hadronic response current, $J^{\mu}$, is defined in the
ejectile plane, defined in Fig.~2 of Ref. \cite{dg89}.
The ALT paper is based on the use of
reduced amplitudes $t_{sm_s\lambda m}$, where
$\lambda$ and $m$ are the virtual photon and deuteron spin projections
in the direction of the momentum transferred by the scattered
electron, ${\bf q}$, and the spins of the outgoing nucleons are coupled
into states
of total spin $s = 0$ or $1$, with total spin projection $m_s$ in the
direction of the  relative momentum ${\bf p}_{np}$ of the outgoing $np$
pair in the center of momentum (c.m.) frame.
For our present task it is a fortunate
coincidence that ALT chose the spin quantization axis for the deuteron to
be in the direction of ${\bf q}$, and the quantization for the final state
nucleon spins to be along the direction of the relative $np$ momentum in
the c.m. frame of the outgoing pair \footnote{This
choice was made in the original treatment\cite{des59}, but was forgotten in
the meantime and that has lead to some confusion. Compare the final state
polarization results in Refs.\cite{ar79} and \cite{alt88}.},
because this makes  it easy
to  identify their spin projections with our helicities, as follows:
\begin{eqnarray}
\lambda&&=\lambda_\gamma \nonumber\\
m&&=-\lambda_D \nonumber\\
m_s&&=\lambda_p -\lambda_n \, . \label{projections}
\end{eqnarray}
However, the ALT decision to work with amplitudes
with a definite value of the total nuclear spin $s$ leads to
subsequent differences in appearance between the two approaches.
In spite of this, the simple relations (\ref{projections})  allow us to
connect our helicity formalism with the ALT formalism
using only Clebsch-Gordan coefficients.

For $s = 1, m_{s} = \pm 1$, the relation to the helicity states is
straightforward:
\begin{mathletters}
\begin{eqnarray}
\label{first}
t_{1 1 \lambda m} &=&
C \langle +{\textstyle{1\over 2}}~-{\textstyle{1\over 2}}\vert J_{\lambda}
\vert -m\rangle \\
t_{1\,-1 \lambda m} &=&
C\langle -{\textstyle{1\over 2}}~+{\textstyle{1\over 2}}\vert J_{\lambda}
\vert -m\rangle ~,
\end{eqnarray}
\end{mathletters}

\noindent where $C$ is a proportionality factor.
\noindent The Jacob-Wick (helicity) parity conservation relation (\ref{parity})
gives the following parity relations for the $m_{s} = \pm 1$ ALT amplitudes
\begin{eqnarray}
t_{1\, \pm 1 \lambda m} &=& C \langle \pm{\textstyle{1\over 2}}~
\mp{\textstyle{1\over 2}}\vert J_{\lambda}
\vert -m\rangle \nonumber \\
&=& (-1)^{m_{s}+\lambda+m} C \langle \mp{\textstyle{1\over 2}}~
\pm{\textstyle{1\over 2}}\vert  J_{- \lambda} \vert
m\rangle \nonumber\\
&=& (-1)^{m_{s}+\lambda+m} t_{1\, \mp1\, -\lambda\, -m} \nonumber\\
&=& (-1)^{1+s+m_{s}+\lambda+m} t_{1\, \mp1\, -\lambda\, -m} \, ,
\end{eqnarray}

\noindent in agreement with Eq.~(4) of ALT.

The comparison of the $s = 1, m_s =0$ and $s = 0, m_s =0$ states is less
straightforward.  In these cases we need to form
the symmetric and antisymmetric normalized linear combinations of the two
outgoing nucleon helicities and find:
\begin{mathletters}
\begin{eqnarray}
\label{second}
t_{1 0 \lambda m} &=&
C {1 \over\sqrt{2}} \left[
\langle+{\textstyle{1\over 2}}~+{\textstyle{1\over 2}}\vert J_{\lambda}
\vert -m\rangle +
\langle-{\textstyle{1\over 2}}~-{\textstyle{1\over 2}}\vert J_{\lambda}
\vert -m\rangle \right]
%\nonumber
\label{triplet} \\
t_{0 0 \lambda m} &=&
C {1 \over\sqrt{2}} \left[
\langle+{\textstyle{1\over 2}}~+{\textstyle{1\over 2}}\vert J_{\lambda}
\vert -m \rangle -
\langle-{\textstyle{1\over 2}}~-{\textstyle{1\over 2}}\vert J_{\lambda}
\vert -m \rangle \right] ~~.
%\nonumber
\label{singlet} \
\end{eqnarray}
\end{mathletters}

\noindent The symmetric combination Eq. (\ref{triplet}) is actually the $s=1$
amplitude because it
has the appropriate phase under the parity transformation as the rest of the
triplet:
\begin{eqnarray}
t_{1 0 \lambda m} &=&
C {1 \over\sqrt{2}} \left[
\langle +{\textstyle{1\over 2}}~
+{\textstyle{1\over 2}}\vert J_{\lambda}
\vert -m\rangle +
\langle-{\textstyle{1\over 2}}~-{\textstyle{1\over 2}}
\vert  J_{\lambda} \vert -m\rangle \right]\nonumber\\
&=& C {1 \over\sqrt{2}} \left[(-1)^{\lambda+m}
\langle -{\textstyle{1\over 2}}~
-{\textstyle{1\over 2}}\vert J_{-\lambda}
\vert m\rangle +
(-1)^{\lambda+m}\langle+{\textstyle{1\over 2}}~+{\textstyle{1\over 2}}
\vert  J_{-\lambda} \vert m\rangle \right]\nonumber\\
&=&(-1)^{\lambda+m} t_{1 0 \, -\lambda\, -m}\nonumber\\
&=&(-1)^{m_{s}+\lambda+m} t_{1 0 \, -\lambda\, -m}\nonumber\\
&=&(-1)^{1+s+m_{s}+\lambda+m} t_{1 0 \, -\lambda\, -m} \, .
\end{eqnarray}

\noindent  The singlet Eq. (\ref{singlet}), on the other hand, is
antisymmetric and hence has the opposite phase under parity:
\begin{eqnarray}
t_{0 0 \lambda m} &=&
C {1 \over\sqrt{2}} \left[
\langle +{\textstyle{1\over 2}}~
+{\textstyle{1\over 2}}\vert J_{\lambda}
\vert -m\rangle -
\langle-{\textstyle{1\over 2}}~-{\textstyle{1\over 2}}
\vert  J_{\lambda} \vert -m\rangle \right]\nonumber\\
&=& C {1 \over\sqrt{2}} \left[(-1)^{\lambda+m}
\langle -{\textstyle{1\over 2}}~
-{\textstyle{1\over 2}}\vert J_{-\lambda}
\vert m\rangle -
(-1)^{\lambda+m}\langle+{\textstyle{1\over 2}}~+{\textstyle{1\over 2}}
\vert  J_{-\lambda} \vert m\rangle \right]\nonumber\\
&=& (-1)^{1+\lambda + m} t_{0 0 \,-\lambda \,-m} \nonumber \\
&=& (-1)^{1+m_{s}+\lambda + m} t_{0 0 \,-\lambda \,-m}
\nonumber \\
&=& (-1)^{1+s+m_{s}+\lambda + m} t_{0 0 \,-\lambda \,-m}~~. \
\end{eqnarray}
All cases can be described by a single
formula
\begin{eqnarray}
t_{s m_{s} \lambda m} &=& (-1)^{1 + s + m_{s} + \lambda + m} t_{s\, -m_{s}\,
 -\lambda \,-m} ~,  \
\end{eqnarray}
which is exactly Eq. (4) of \cite{ar93}. Thus we have shown that
the ALT amplitudes are simple linear combinations of the helicity
amplitudes.  From
this point on the comparison between the two papers is strictly a matter
of transcription of the results from one notation to the other.

Of course, in
matters like this, transparent and concise notation is very important,
and following a
suggestion by Moravcsik, we found that a hybrid transversity basis gave
comparatively simple results.  Our hybrid basis is obtained from the
helicity basis by rotating the amplitudes by $-\pi/2$ around the $x$ axis.
This is equivalent to using amplitudes in which hadron spins are quantized
with respect to the $y$ axis, but the photon spin remains quantized with
respect to the $z$ axis.  The construction of this basis is described in
Sec.~II.E of DG, and the explicit transformations for helicity basis to hybrid
basis are given in the Appendix of that article.  This basis could also be
expressed in terms of the ALT amplitudes by combining the relations (3) and
(5) with the transformations given in DG.  Final results for the
observables in which $N_1=p$, expressed as bilinear products of the hybrid
amplitudes, are summarized in Tables  X--XII of DG.   Because of the use
of the hybrid basis, one-half of the entries in these tables are zero, and
{\it all} of the 162 nonzero entries are linear combinations of 162 {\it
different\/} real bilinear products.  In the language of ALT, there are
{\it no\/} linear  relations connecting these products to each other.

As discussed in DG, it is possible to obtain a simple understanding of the
origin of the 162 independent real bilinear products determined by reactions
in which $N_1=p$.  These measurements divide the 18 independent complex
amplitudes into two disjoint classes of
9 amplitudes each [defined explicitly in Eq.~(99) of DG], in the sense that
these measurements determine all products of amplitudes in each class,  but
no products of amplitudes in one class with those in another.  Hence
$N_1=p$ measurements determine $9\times9+9\times9=162$ independent products.
Now, the same Tables X--XII can be used to obtain the observables for
neutron measurements (in which $N_1=n$), provided one {\it exchanges five of
the amplitudes in one class with five in the other\/} (see DG for details).
Hence neutron measurements determine $2\times(4\times5+5\times4)=80$ new
real bilinear products, the remaining 82 being identical to those already
fixed by the proton measurements. In the ALT language, the identity of the
82 products which occur in both proton and neutron measurements could be
written as linear relations, but the simplicity of the pattern of results
given in Tables X--XII makes this unnecessary.  Finally, the remaining
$2\times(4\times4+5\times5)=82$ products of amplitudes arising from
products amplitudes with one from each  of the two groups of 5 amplitudes
exchanged in the $p\to n$ substitution, and similar products between the two
groups of 4 amplitudes not exchanged in the $p\to n$ substitution, cannot be
determined by either class of experiments, and require
$d(e,e'\,\vec{p}\,\vec{n})$ experiments, as stated at the beginning of this
comment. That does {\it not} mean, however, that these experiments are
necessary for the complete separation of amplitudes.

We conclude by emphasizing that any attempt to discuss complete separations
in reactions as complex as deuteron electrodisintegration, or to find ``the
most suitable complete set'' of amplitudes, requires that the relations
between the observables and the bilinear products from which they are
determined be as simple as possible.  We believe that the hybrid transversity
basis, popularized by Moravscik and developed in DG, is just such a basis.
The ALT choice of the final state spin quantization axis makes a
simple, direct link between their amplitudes and the hybrid basis possible.
Despite the fact that one can view the problem of complete separation of
amplitudes as solved by this basis, it would still be satisfying to see
the ALT results expressed in this basis.

\begin{acknowledgements}
The support of the US Department of Energy under Grant Nos.~DE-FG03-93DR40774
(VD)
and DE-FG05-88ER40435 (FG) is gratefully acknowledged.
\end{acknowledgements}

 \end{document}